\documentclass[sigconf]{acmart}

\usepackage{multirow}
\usepackage[ruled,linesnumbered]{algorithm2e}

\copyrightyear{2020}
\acmYear{2020}
\setcopyright{iw3c2w3}
\acmConference[WWW '20]{Proceedings of The Web Conference 2020}{April 20--24, 2020}{Taipei, Taiwan} \acmBooktitle{Proceedings of The Web Conference 2020 (WWW '20), April 20--24, 2020, Taipei, Taiwan} \acmPrice{}
\acmDOI{10.1145/3366423.3380295} \acmISBN{978-1-4503-7023-3/20/04}

\begin{document}

%% allowing the author to define a "short title" to be used in page headers.
\title{Leveraging Code Generation to Improve Code Retrieval and Summarization via Dual Learning}

%%
%% The "author" command and its associated commands are used to define
%% the authors and their affiliations.
%% Of note is the shared affiliation of the first two authors, and the
%% "authornote" and "authornotemark" commands
%% used to denote shared contribution to the research.
\author{
	Wei Ye$^{1 \dagger}$, Rui Xie$^{1,2 \dagger}$
}
\author{
Jinglei Zhang$^{1}$, Tianxiang Hu$^{1,2}$, Xiaoyin Wang$^{3}$, and Shikun Zhang$^{1 \ddagger}$
} 

\thanks{$\dagger$ Both authors contributed equally to this research.}
\thanks{$\ddagger$ Corresponding author.}

\affiliation{
	$^1$ National Engineering Research Center for Software Engineering, Peking University, Beijing, China
}
\affiliation{
	$^2$School of Software and Microelectronics, Peking University, Beijing, China
}
\affiliation{
	$^3$Department of Computer Science, University of Texas at San Antonio, San Antonio, USA
}

\email{{wye, ruixie, jinglei.zhang, hutianyang}@pku.edu.cn}
\email{xiaoyin.wang@utsa.edu, zhangsk@pku.edu.cn}

%%
%% By default, the full list of authors will be used in the page
%% headers. Often, this list is too long, and will overlap
%% other information printed in the page headers. This command allows
%% the author to define a more concise list
%% of authors' names for this purpose.
\renewcommand{\shortauthors}{Wei Ye and Rui Xie, et al.}

%%
%% The abstract is a short summary of the work to be presented in the
%% article.
\begin{abstract}
Code summarization generates brief natural language description given a source code snippet, while code retrieval fetches relevant source code given a natural language query. Since both tasks aim to model the association between natural language and programming language, recent studies have combined these two tasks to improve their performance. However, researchers have yet been able to effectively leverage the intrinsic connection between the two tasks as they train these tasks in a separate or pipeline manner, which means their performance can not be well balanced. In this paper, we propose a novel end-to-end model for the two tasks by introducing an additional code generation task. More specifically, we explicitly exploit the probabilistic correlation between code summarization and code generation with dual learning, and utilize the two encoders for code summarization and code generation to train the code retrieval task via multi-task learning. We have carried out extensive experiments on an existing dataset of SQL and Python, and results show that our model can significantly improve the results of the code retrieval task over the-state-of-art models, as well as achieve competitive performance in terms of BLEU score for the code summarization task.
\end{abstract}

%%
%% The code below is generated by the tool at http://dl.acm.org/ccs.cfm.
%% Please copy and paste the code instead of the example below.
%%

%%
%% Keywords. The author(s) should pick words that accurately describe
%% the work being presented. Separate the keywords with commas.

%%
%% This command processes the author and affiliation and title
%% information and builds the first part of the formatted document.
\maketitle

\section{Introduction}\label{sec:introduction}

Modern software engineering relies heavily on large amount of third-party libraries and public codes from various websites. During software development and maintenance, developers often spend lots of time understanding code and searching for the code snippets they need~\cite{AllamanisBDS18}. Therefore, code retrieval and code summarization play an important role in many software engineering activities. Code summarization generates brief natural language description given a source code snippet, while code retrieval fetches relevant source code given a natural language query. As for both academics and industry of software engineering, they are two important and challenging tasks.

A large amount of code and related text information have been accumulated on the Web. For example, Stack Overflow ~\cite{SO} contributes to a huge amount of code snippets, usually paired with natural-language-based questions and comments. Researchers have extracted <natural language description, code> pairs from those resources to help develop data-hungry models for associating natural languages with programming languages~\cite{CodeNN,StaQC}. For example, StaQC~\cite{StaQC} is a large-scale dataset automatically mined from Stack Overflow, which contains more than 100K <question, code> pairs for SQL and Python respectively. Iyer el al.~\cite{CodeNN}  built a dataset for SQL and C\# in a similar way. These datasets have greatly contributed to the progress of research on code retrieval and code summarization.

With these datasets mined from the Web, deep learning is now widely used in code retrieval and code summarization as a mainstream approach. Researchers in the fields of Web technologies, natural language processing, deep learning and software engineering have proposed a variety of neural models for these two tasks, aiming to design better features or more sophisticated network structures to capture more accurate semantics of code and natural language text. For example, the code retrieval model DCS~\cite{DCS} used two neural encoders to model the semantics of natural language queries and code snippets respectively, and measured their correlation by cosine similarity of the encoders' output. There are more deep learning works on code summarization due to its resemblance to machine translation. In particular, SBT~\cite{SBT} transformed an abstract syntax tree (AST) into a token sequence to generate a code summary, in which brackets were used to represent the tree structure. Code2Seq~\cite{code2seq} represented a code snippet as a set of compositional paths in its AST. Wan et al.~\cite{Yao2018Improving} adopted Tree-RNN and reinforcement learning to improve code summarization.

Since both code retrieval and code summarization tasks aim to model the association between natural language and programming language, recent studies have combined these two tasks to improve their performance. Chen et al.~\cite{BVAE} proposed BVAE, a neural framework that contains two Variational-Auto-Encoders (VAEs) to model source code and natural language respectively. Both VAEs are trained jointly to reconstruct their inputs with regularization that captures the closeness between the latent variables of code and description. CoaCor~\cite{CoaCor} trained a code summarization model to generate code summaries that can be used for the retrieval task based on reinforcement learning. These approaches, however, have yet effectively leveraged the intrinsic connection between the two tasks. For example, the BVAE models for retrieval and summarization tasks are designed and trained independently. CoaCor adopted a pipeline approach to feed generated summary to a separate assemble module for code retrieval. These simplified solutions would lead to two problems:

\begin{itemize}
\item Performance cannot be well balanced between code retrieval and code summarization. For example, the BLEU score~\cite{BLEU} of the code summarization model in CoaCor is not satisfactory though it improves the performance of existing code retrieval models significantly. Different from what claimed in ~\cite{CoaCor}, we respectively argue that generating summaries close to human-provided queries is naturally fit to code retrieval. The compromise of BLEU score, which represents the similarity between the generated summaries and human-written ones, can be avoided if we can model the inner connection between the two tasks better.

\item The complexity of overall procedure makes the model training more challenging. For example, BVAE~\cite{BVAE} only provides a proof-of-concept implementation based on simplest network architecture, because more powerful models are harder to train when VAE is applied to text. Similarly, the convergence of reinforcement learning in CoaCor~\cite{CoaCor} is also a problem.

\end{itemize}

To overcome these issues, we propose an easy-to-train, end-to-end method to improve both code retrieval and code summarization. Our work is inspired by CoaCor~\cite{CoaCor}. If the similarity between generated summaries and code-search queries can help improve code retrieval, it is natural to assume that the similarity between generated code and code snippets in code bases can also be beneficial. Therefore, we introduce an additional code generation task. Since code summarization and code generation are two tasks that have dual forms, we explicitly exploit the probabilistic correlation between code summarization and code generation with dual learning~\cite{DSL}. The two encoders for code summarization and code generation are also shared with a code retrieval scorer via multi-task learning. In this way, we get an extremely simple yet effective model, named \textit{CO3}, in which the \textbf{3} \textbf{CO}de-related tasks can be trained simultaneously.

In this paper, we make the following contributions:

\begin{itemize}

\item We design a simple yet effective end-to-end model for both code retrieval and code summarization by introducing the code generation task and exploiting the intrinsic connection between these tasks via dual learning and multi-task learning. 

\item We carried out extensive experiments on an existing dataset of SQL and Python. Experiment results show that our model can improve code retrieval significantly compared to state-of-the-art models, without affecting the performance in code summarization.

\item With ablation study and case study, we provide strong evidence that the introduction of code generation and dual learning leads to better representations of source code and text in terms of semantics.

\end{itemize}

The rest of this paper is organized as follows. In Section 2, we introduce some background knowledge on code retrieval, code summarization, code generation and dual learning. In Section 3, we explain our approach in details. In Section 4, we describe our experiment setup, including datasets and evaluation metrics used. In Section 5, we present and analyze our experiment results. Before we conclude in Section 7, we discuss related research efforts in the areas of code retrieval, code summarization, dual learning and multi-task learning.  

\section{Background}

\subsection{Code Retrieval}
Code retrieval aims to match code with natural language query. Formally, given a set of code snippets $X$ and a natural language query $y$, code retrieval aims to retrieve the corresponding code snippet $x \in X$ that matches the semantics of the query $y$. Firstly, we have an encoder to represent the code snippet $x$ as code vector $V_x$, which is calculated as follows:

\begin{equation}
    V_x = E_x(x)
\end{equation}
where $E_x(\cdot)$ is the neural network to encode source code. Then, we map the natural language query $y$ to the same semantic space as code vector $V_x$ , represented as query vector $V_y$, which is calculated as follows:

\begin{equation}
    V_y = E_y(y)
\end{equation}
where $E_y(\cdot)$ is another neural network to encode natural language queries Finally, the similarity value in the matching step is calculated as follow:

\begin{equation}
    Similarity = S(V_x, V_y)
\end{equation}
where $S(\cdot)$ is a similarity function between code vector $V_x$ and query vector $V_y$. By maximizing the similarity function, we can get the most relative code to a given description. 

\subsection{Code Summarization and Code Generation}
As we mentioned above, code summarization could be treated as a text generation task. Given an input code snippet $x = (x_1, x_2, ..., x_m)$ that has $m$ code tokens, code summarization aims to generate a readable natural language query $y = (y_1, y_2, ..., y_n)$ with $n$ words which describe the input code snippet $x$. Let $Y$ be the set of all possible summary sequences. The system tries to find the optimal sequence $y$ for $x$:
\begin{equation}
    \arg \max_{y \in Y} P(y|x)
\end{equation}

In contrast, let $X$ be the set of all possible code snippets, given a natural language query $y$, code generation is to generate the optimal code snippets $x$ for $y$:
\begin{equation}
    \arg \max_{x \in X} p(x|y)
\end{equation}

Note that for these two Seq2Seq models, the input of code summarization is the expected result of code generation, and vice versa for code generation. Thus, as we mentioned in Section 1, code summarization and code generation are dual tasks of each other.

\subsection{Dual Learning}
Dual Learning is introduced in~\cite{DSL}, which aims to utilize the duality so that two dual tasks can keep learning from each other until convergence. Given two tasks: a primal task that takes samples from space $\mathcal{X}$ as input to map into output space $\mathcal{Y}$, and a dual task that takes samples from space $\mathcal{Y}$ as input to map to output space $\mathcal{X}$. Then, for $n$ training pairs ${(X_i,Y_i)}^n_{i=1}$, the primal task is to find function $f: \mathcal{X} \mapsto \mathcal{Y}$, and the dual task is to find function $g: \mathcal{Y} \mapsto \mathcal{X}$.

With the principle of duality, if the learned primal and dual models are perfect, we should have:
\begin{equation}
 P(x)P(y|x; \theta_{xy}) =P(x,y)= P(y)P(y|x; \theta_{yx}) , \forall x,y 
\end{equation}
where $\theta_{xy}$ is the learned parameter from $f$, $\theta_{yx}$ is the learned parameter from $g$, and we call this property probabilistic duality. 
 
By incorporating dual learning into specific training objectives of deep learning models, we would have:

\begin{equation}
    \mathrm{Objective 1: } \min \limits_{\theta_{xy}} \sum_{i=1}^{|\mathcal{X}|}\mathcal{L}_{xy}(f(X_i; \theta_{xy}), Y_i)
\end{equation}
\begin{equation}
    \mathrm{Objective 2: } \min \limits_{\theta_{yx}} \sum_{i=1}^{|\mathcal{Y}|}\mathcal{L}_{yx}(g(Y_i; \theta_{yx}), X_i)
\end{equation}
\begin{equation}
    \mathrm{s.t. } P(X_i)P(Y_i|X_i; \theta_{xy}) = P(Y_i)P(X_i|Y_i; \theta_{yx}), \forall i \in [1, |\mathcal{X}|]
\end{equation}
where $\mathcal{L}_{xy}$ and $\mathcal{L}_{yx}$ are the loss functions decided by the function $f$ and $g$. $X$ and $Y$ are the sets of all training pairs.

To optimize these training objectives, the common practice in dual learning is to introduce Lagrange multipliers and add the equality constraint of probabilistic duality into the objective functions. Convert the probabilistic duality constraint into the following regularization term:

\begin{equation}
\begin{aligned}
    \mathcal{L}_{dual}&=(\exp(P(x))+\exp(P(y|x, \theta_{xy}))\\
    &-exp(P(y))-exp(P(x|y, \theta_{yx})))^2
\end{aligned}
\end{equation}
and the training objectives become:

\begin{equation}
\begin{aligned}
    \mathrm{Objective 1': } \min \limits_{\theta_{xy}} \sum_{i=1}^{|\mathcal{X}|}[&\mathcal{L}_{xy}(f(X_i; \theta_{xy}), Y_i) \\
    &+ \mathcal{L}_{dual}(X_i, Y_i; \theta_{xy}, \theta_{yx})]
\end{aligned}
\end{equation}
\begin{equation}
\begin{aligned}
    \mathrm{Objective 2': } \min \limits_{\theta_{yx}} \sum_{i=1}^{|\mathcal{Y}|}[&\mathcal{L}_{yx}(g(Y_i; \theta_{yx}), X_i) \\
    &+ \mathcal{L}_{dual}(X_i, Y_i; \theta_{xy}, \theta_{yx})]
\end{aligned}
\end{equation}
which can be trained using common optimization methods. In this paper, we utilize this regularization term to restrain code summarization and code generation to achieve a better performance.

\section{Approach}

\subsection{Overview}

In this section, we formulate the problem and describe our model of CO3.

We take the dual learning mechanism in ~\cite{DSL} and propose two dual tasks: a primal code summarization task that takes source code sequence $x$ as input and summarizes it into text sequence $y'$; and a dual code generation task that takes text sequence $y$ as input and uses it to generate code sequence $x'$. We reuse $x$ and $y$ to supervise $x'$ and $y'$ respectively, with dual learning mechanism to improve the performance of both tasks. Afterwards, we use the hidden states of these two tasks to facilitate and improve the performance of code retrieval task. 

\begin{figure}[ht]
  \centering
  \includegraphics[width=\linewidth]{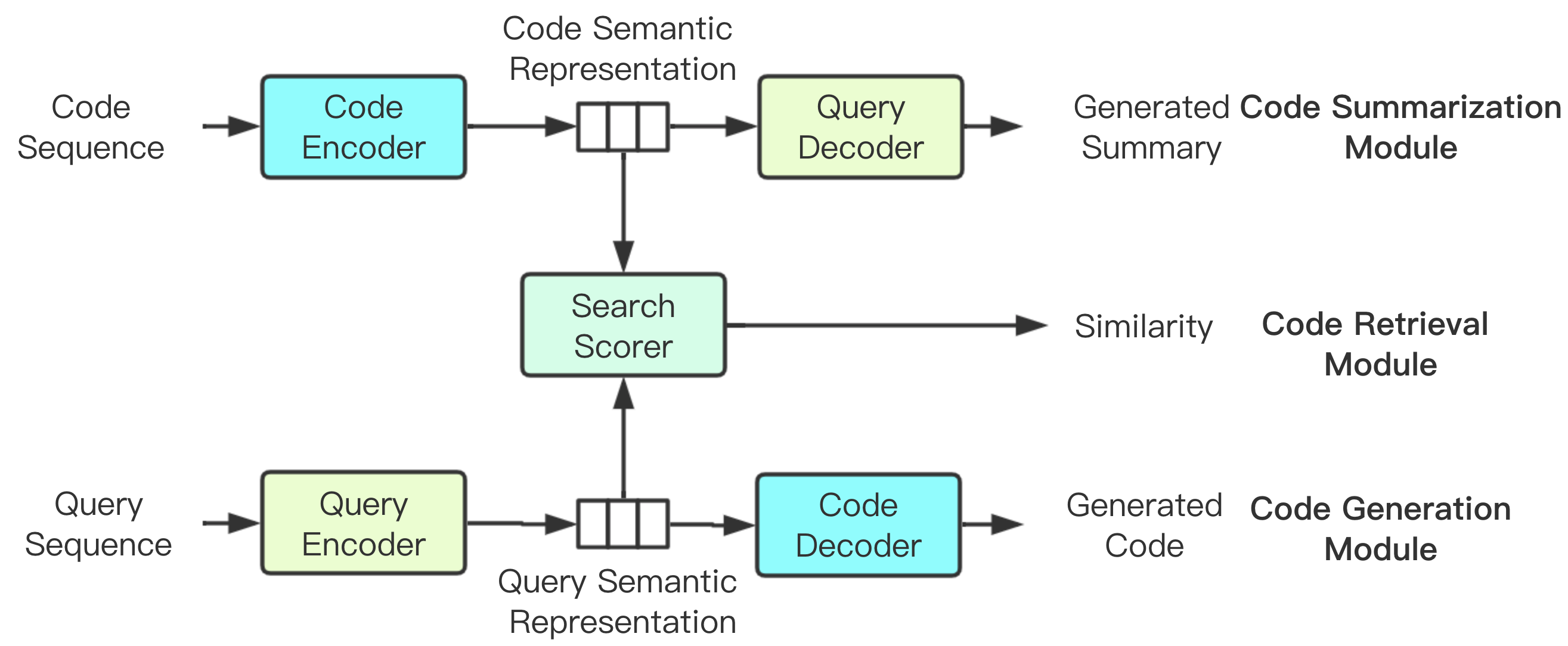}
  \caption{Overall framework of CO3. The blocks with the same color indicate that these modules use same LSTM-cell parameters. More specifically, Code Encoder and Code Decoder share the same LSTM-cell instance; Query Encoder and Query Decoder share the same LSTM-cell instance.}
  \label{fig:framework}
\end{figure}

% \begin{figure*}[ht]
%   \centering
%   \includegraphics[width=0.8\linewidth]{images/s2s.png}
%   \caption{Sequence-to-sequence structure of the code summarization module, which takes source code as input and then outputs its text summary. The code generation module is similar to this, with the inputs and embedding layers switched.}
%   \label{fig:s2s}
% \end{figure*}

The model is split into three parts, as shown in Figure \ref{fig:framework}: 

\begin{enumerate}
    \item The code summarization module encodes the code sequence $x$, and summarizes it into a text sequence $y'$.
    \item The code generation module encodes the text sequence $y$, and uses it to generate a code sequence $x'$.
    \item The code retrieval module calculates similarity scores between the hidden states of the code summarization module and the code generation module, and then retrieves matching source code based on the scores.
\end{enumerate}

Our model has two major features:

\begin{enumerate}
    \item We add a restriction between the code summarization module and the code generation module with dual learning method to connect them and help capture more intrinsic and precise representations of text and code.
    
    \item We share the parameters of LSTM cells between the encoder of the code summarization module and the decoder of the code generation module, since they both deal with source code; so do the decoder of the code summarization module and the encoder of the code generation module. This reduces the number of model parameters. In this way, we use only two LSTM instances for the three tasks, constructing an extremely simple model.
    
\end{enumerate}

\subsection{Embedding Layer}

To deal with both natural language texts and source code snippets, we use separate embedding layers to convert input sequences into high-dimensional vectors. 

For source code input $x$, we directly use code snippets processed by StaQC~\cite{StaQC}, which is a code sequence $\{x_1, x_2, ..., x_m\}$. Then we use an embedding matrix $E_x$ to map the tokens in code sequence into a high-dimensional vector space to get the code input embeddings $\{ E_x(x_1), E_x(x_2), ..., E_x(x_m)\}$.

For natural language input $y$, we simply split the sequence by space and covert them into one-hot representations. Then we use an embedding matrix $E_y$ to map the one-hot representation of each word into a high-dimensional vector space to get the text input embeddings $\{E_y(y_1), E_y(y_2), ..., E_y(y_n)\}$.

\subsection{Code Summarization Module}

To begin with, we employ a Bi-Directional Long Short-Term Memory Network(Bi-LSTM)-based Recurrent
Neural Network (RNN)~\cite{LSTM} encoder on the source code embedding sequence $\{E_x(x_t)\}$ to model the temporal interactions between words.
% , as shown in Figure~\ref{fig:s2s}.

The LSTM cell is composed of three multiplicative gates. At each time step $t$, it takes the embedding $E_x(x_t)$ of the $t^{th}$ word in the source code sequence, and then merges it with the last hidden state $h^e_{t-1}$ to generate a new hidden state. The transmission rate of each input is calculated by the gates in LSTM cell. At each time step $t$, the hidden state is updated as:

\begin{equation}
    i=\sigma(W_ih^e_{t-1}+U_iE_x(x_t)+b_i)    
\end{equation}
\begin{equation}
    f=\sigma(W_fh^e_{t-1}+U_fE_x(x_t)+b_f)    
\end{equation}
\begin{equation}
    o=\sigma(W_oh^e_{t-1}+U_oE_x(x_t)+b_o)    
\end{equation}
\begin{equation}
    g=\tanh(W_gh^e_{t-1}+U_gE_x(x_t)+b_g)    
\end{equation}
\begin{equation}
    g_t=f\odot g_{t-1}+i \odot g    
\end{equation}
\begin{equation}
    h_t=o\odot \tanh(g_t)    
\end{equation}
where $\sigma$ is the element-wise sigmoid function and $\odot$ is the element-wise product; $W_i,W_f,W_o,W_g$ are weight matrices for last hidden states; $U_i,U_f,U_o,U_g$ are weight matrices for source code embeddings; $b_i, b_f, b_o, b_g$ are biases. For simplicity, we denote the above calculation as below (the memory cell vector $g_{t-1}$ is omitted):

\begin{equation}
  h^e_t = \mathrm{LSTM}_x(h^e_{t-1}), E_x(x_t))
\end{equation} where $h^e_t$ denotes the hidden state of the $t^{th}$ step in Bi-LSTM encoder for $x$. Since $h^e_t$ only receives information from codes before position $t$, we use Bi-Directional LSTM to further incorporate information after position $t$ into hidden states:

\begin{equation}
    \overrightarrow{h}^e_t = \overrightarrow{\mathrm{LSTM}}_x (\overrightarrow{h}^e_{t-1}), E_x(x_t))
\end{equation}
\begin{equation}
    \overleftarrow{h}^e_t = \overleftarrow{\mathrm{LSTM}}_x (\overleftarrow{h}^e_{t+1}), E_x(x_t))
\end{equation}
\begin{equation}
    h^e_t = [\overrightarrow{h}^e_t; \overleftarrow{h}^e_t]
\end{equation}

Then we employ another LSTM-based decoder to generate text summary:

\begin{equation}
  h^d_t = \mathrm{LSTM}_y(h^d_{t-1}, [E_y(y_{t-1}') ; c^d_{t-1}])
\end{equation}
where $h^d_t$ denotes the hidden state of the $t^{th}$ step in LSTM decoder for $h^e$, $h^d_0 = [\overrightarrow{h}^e_m;\overleftarrow{h}^e_1]$; $y_{t-1}'$ denotes the generated word of the $t-1^{th}$ step, ``;`` denotes the concatenation operation, and $c^d_{t-1}$ is the context vector produced by the standard attention mechanism:

\begin{equation}
    \delta_{it}=\frac{\exp(f(h^e_i, h^d_t))}{\sum_{i=1}^m\exp(f(h^e_i, h^d_t))}
\end{equation}
\begin{equation}
    c_t=\sum_{i=1}^m\delta_{it}h^e_i
\end{equation}
where $f$ is a trainable function to calculate the similarity between hidden states. We use a simple bi-linear function $f(x,y)=xW_{bi}y$ where $W_{bi}$ is a trainable matrix.

Finally, the context vector $c_t$ is concatenated with the decoder output $h^d_t$ and fed into a linear layer to obtain the generated word distribution $P_v$:

\begin{equation}
    u^e_t = W_u[h^d_t;c_t] + b_u    
\end{equation}
\begin{equation}
    P_v = softmax(W_v u^e_t + b_v)   
\end{equation}

The objective is to maximize the probability of generating $y$ with input $x$, so the loss function here is designed to be the negative log likelihood of target word $y_t$:

\begin{equation}
    \mathcal{L}_{cs}(x, y, \theta_{cs}) = -\sum\log(P_v(y_t))    
\end{equation}
where $\theta_{cs}$ denotes all the trainable parameters in the code summarization module.

\subsection{Code Generation module}

To help our primal task---code summarization---achieve better performance, we construct a dual task with an opposite network, so it would output the source code sequence from its summary.

The whole structure is basically the same as the code summarization module, so we simply denote it as:

\begin{equation}
  \hat h^e_t = \mathrm{BiLSTM}_y(\hat h^e_{t-1}, E_y(y_t))
\end{equation}
\begin{equation}
  \hat h^d_t = \mathrm{LSTM}_x(\hat h^d_{t-1}, [E_x(x_{t-1}') ; \hat c_{t-1}^{d}])
\end{equation}
where $\hat h^e_t$ and $\hat h^d_t$ denote the hidden states of the $t^{th}$ step in the Bi-LSTM encoder and the LSTM decoder, respectively; $x_{t}'$ denotes the generated word of the $t^{th}$ step; and $\hat c^d_{t-1}$ denotes the context vector.

Since both the code generation encoder and the code summarization decoder try to work with source code input, we believe sharing the LSTM-cell parameters between them would reduce the model's complexity. The same is true for the code generation decoder and the code summarization encoder. We find that performance does not change much when using or not using individual encoders/decoders.

The loss function is also designed to be the negative log likelihood of target word $y_t$:

\begin{equation}
    \mathcal{L}_{cg}(x, y, \theta_{cg}) = -\sum\log(\hat P_v(x_t))    
\end{equation}
where $\hat P_v$ is the generated word distribution calculated by $\hat h^d_t$ and $\hat c_{t-1}$, and $\theta_{cg}$ denotes all the trainable parameters in the code generation module.

\subsection{Dual Learning for Code Summarization and Generation}

Referring to the idea of dual learning in ~\cite{DSL}, we hope that the code summarization module and the code generation module reflect the same data distributions. So it is reasonable to restrict these two modules by the following equation:

\begin{equation}
    P(x) P(y|x)=P(x,y)=P(y) P(x|y)
\end{equation}

In the equation, $P(y|x)$ and $P(x|y)$ can be calculated by the code summarization module and the code generation module. But $P(x)$ and $P(y)$ are marginal distributions which cannot be directly calculated, so we use language models to fit their real values. We pre-train these language models with input corpus of natural language and source code separately. And the marginal distribution of a sentence $s$ can be defined as:

\begin{equation}
    \Pi^{|S|}_{i=1} P(s_i|s_{<i})
\end{equation}
where $P(s_i|s_{<i})$ is calculated by the pre-trained language model.

\subsection{Code Retrieval Module}

For code retrieval task with natural language input $\tilde y$, we first calculate the similarities between $\tilde y$ and all candidate source code $\{\tilde x\}$, and then choose those code sequences with highest scores as output.

The similarity score is defined as :

\begin{equation}
    \mathrm{Score}(\tilde y, \tilde x)=f_{cr}(h^e, \hat h^e)
\end{equation}
where $h^e$, $\hat h^e$ is calculated by the encoders' hidden states in the code summarization module and the code generation module when they take $\tilde x$ and $\tilde y$ as input:

\begin{equation}
    h^e = \tanh(maxpooling(h^e_1, h^e_2, ..., h^e_m))
\end{equation}
\begin{equation}
    \hat h^e = \tanh(maxpooling(\hat h^e_1, \hat h^e_2, ..., \hat h^e_m))
\end{equation}
and $f_{cr}$ is the similarity function of vectors:

\begin{equation}
    f_{cr}(p, q)=\mathrm{cos}(W_p p + b_p, W_q q + b_q)
\end{equation}
where $\mathrm{cos}$ denotes cosine similarity, and $W_p, b_p, W_q, b_q$ are trainable parameters.

Then we use Ranking Loss here to define the training objective: for a paired sample $(x, y)$, we randomly choose another source code input $x^r$ and a text summary $y^r$, and we assume the score of pair $(x, y)$ is higher than that of pair $(x^r, y^r)$ with a margin of at least $M_{cr}$. So the loss function is:

\begin{equation}
    \mathcal{L}_{cr}(x, y, x^r, y^r, \theta_{cr}) = \max(f_{cr}(x, y)-f_{cr}(x^r, y^r)-M_{cr},0)
\end{equation}
where $\theta_{cr}$ denotes all the trainable parameters in the code retrieval module.

\subsection{Training}

The final training objectives are:
\begin{equation}
    \mathrm{1. } \qquad \qquad \min \limits_{\theta_{cs}} \sum_{i=1}^{|X|} \mathcal{L}_{cs}(X_i, Y_i, \theta_{cs})
\end{equation}
\begin{equation}
    \mathrm{2. } \qquad \qquad \;\; \min \limits_{\theta_{cg}} \sum_{i=1}^{|X|} \mathcal{L}_{cg}(X_i, Y_i, \theta_{cg})
\end{equation}
\begin{equation}
    \mathrm{3. \quad\;} \min \limits_{\theta_{cr}} \sum_{i=1}^{|X|} \mathcal{L}_{cr}(X_i, Y_i, X^R_i, Y^R_i, \theta_{cr})
\end{equation}
\begin{equation}
    \mathrm{s.t.\;\;} P(x)P(y|x, \theta_{cs})=P(y)(P(x|y, \theta_{cg})
\end{equation}
where $x^r$ and $y^r$ in $X^R$ and $Y^R$ are randomly sampled from $X$ and $Y$.

To optimize the training objectives, following the common practice in dual learning, we use Lagrange multipliers and add the equality constraint of probabilistic duality into the objective functions, by converting the probabilistic duality constraint into the following regularization term:

\begin{equation}
\begin{aligned}
    \mathcal{L}_{dual}&=(\exp(P(x))+\exp(P(y|x, \theta_{cs}))\\
    &-exp(P(y))-exp(P(x|y, \theta_{cg})))^2
\end{aligned}
\end{equation}
where $P(x)$ and $P(y)$ are calculated by pre-trained language models; $P(y|x)$ and $P(x|y)$ are calculated by the code summarization module and the code generation module.

Then we train the models by minimizing the weighted combination of the original loss functions and the added regularization term. The algorithm is shown in Algorithm 1.

\begin{algorithm}[t]
\KwData{Marginal distributions $\hat P(X_i)$ and $\hat P(Y_i)$ for any $i \in [1, |X|]$; Lagrange parameters $\lambda_{cs}$ and $\lambda_{cg}$; optimizers $Opt_1$, $Opt_2$, $Opt_3$;}
\KwResult{Trained parameters $\theta_{cs}$, $\theta_{cg}$ and $\theta_{cr}$}
Initialization\;
\While{model not converged}{
  Get a mini-batch of $m$ pairs $\{(X_j, Y_j)\}^m_{j=1}$\;
  Calculate the gradients of the code summarization module:\\
  $\begin{aligned}
  G_{cs} &= \nabla_{\theta_{cs}}\frac{1}{m}\sum_{j=1}^m[\mathcal{L}_{cs}(X_j, Y_j;\theta_{cs})\\
  &+ \lambda_{cs}\mathcal{L}_{dual}(X_j, Y_j;\theta_{cs},\theta_{cg})]
  \end{aligned}$;

  Update the parameters of the code summarization module: \\
  $\theta_{cs} \leftarrow Opt_1(\theta_{cs}, G_{cs})$;

  Calculate the gradients of the code generation module:\\
  $\begin{aligned}
      G_{cg} &= \nabla_{\theta_{cg}}\frac{1}{m}\sum_{j=1}^m[\mathcal{L}_{cg}(Y_j, X_j;\theta_{cg})\\
      &+ \lambda_{cg}\mathcal{L}_{dual}(X_j, Y_j;\theta_{cs},\theta_{cg})]
  \end{aligned}$;

  Update the parameters of the code generation module:\\
  $\theta_{cg} \leftarrow Opt_2(\theta_{cg}, G_{cg})$;

  Randomly sample m ($X^r$, $Y^r$) from $\hat P(X_i)$ and $\hat P(Y_i)$;
  
  Calculate the gradients of the code retrieval module:\\
  $\begin{aligned}
      G_{cr} = \nabla_{\theta_{cr}}\frac{1}{m}\sum_{j=1}^m\mathcal{L}_{cr}(X_j, Y_j, X^R_j, Y^R_j;\theta_{cr})
  \end{aligned}$;

  Update the parameters of code retrieval module:\\
  $\theta_{cr} \leftarrow Opt_3(\theta_{cr}, G_{cr})$;
  
}

\caption{Supervised Learning Algorithm of CO3}
\end{algorithm}

\section{Experiments}

\subsection{Details of Datasets}

To ensure the generality of our experiment results, we used the existing dataset StaQC~\cite{StaQC}, which contains data of two different programming languages---SQL and python---to evaluate our approach.

StaQC was provided by Yao et al.~\cite{StaQC}, which is the largest dataset in SQL and Python domain. It contains 119,519 and 147,546 <question title, code> pairs in SQL and Python respectively, which were mined from Stack Overflow. We consider the question title as the query text of the code snippets in code retrieval task, and in code summarization task it is treated as the corresponding text summary for code. We followed the division of dataset in ~\cite{StaQC}, and used 75\% of the pairs for training, 10\% for validation and 15\% for testing.

\subsection{Implementation Details}

We set the dimensionality of the LSTM hidden states, code embeddings, query embeddings to 400, 200, 200, respectively, following CoaCor~\cite{CoaCor}.  Based on our observation of query text and code text, we set the maximum lengths for query sequence and code sequence as 200 and 120, respectively. A small, fixed $M_{cr}$ value of 0.05 was used in all experiments. To evaluate the code retrieval performance, we randomly selected another 49 code snippets for each <query, code snippet> pair, which were sorted according to the similarity score calculated with the query and the corresponding code snippet. Next we chose the model with the highest validation performance, and computed evaluation metrics on the test set. Adam was used for parameter optimization and the learning rate was set to 0.001. We implemented our model in Pytorch, and trained our models on Tesla T4. We will open-source our code in the near future.

In our experiments, training with dual learning (65 min/epoch) took more time than training without dual learning (20 min/epoch), due to the calculation of the regularization term of duality. The speed of convergence was similar (both averagely 4 epochs). Note that the training overhead due to dual learning is a one-time cost, and the inference speed is not affected.

\subsection{Evaluation Metrics}

\subsubsection{Code Retrieval}

Following previous works, we evaluated the retrieval performance of CO3 and baselines based on MRR~\cite{MRR} and NDCG~\cite{NDCG} metrics, which are widely used in evaluating performance of code retrieval tasks. MRR is a popular metric used for evaluating ranking result. It calculates the Mean Reciprocal Rank over the entire set, rewarding each item with the reciprocal of its order. A higher value of MRR value indicates better performance of code retrieval. NDCG is also widely used in evaluating rankings. It adds the score of an item to each item after it, sums all scores up, and normalizes the scores to the range [0, 1]. It is similar to MRR but with different distribution of weights on each item. Here we choose relevance weight as 1 if the corresponding code snippet is positive and 0 otherwise.

\subsubsection{Code Summarization}

Following previous works, we evaluated the summarization performance of
CO3 and baselines based on BLEU~\cite{BLEU} and METEOR~\cite{METEOR} (Banerjee and Lavie 2005) metrics, which are widely used in evaluating performance of text generation tasks. BLEU score is a popular accuracy-based metric for Neural Machine Translation, and is also used in the code summary generation task. It calculates the similarity between the generated sequence and reference sequence by counting the n-grams that appear in both the candidate sequence and the reference sequence. METEOR measure is the harmonic average of precision and recall, and a prior study~\cite{METEOR} shows that recall-based metrics can be more correlative to manual judgement than accuracy-based metrics like BLEU.

\subsection{Baselines}

\subsubsection{Code Retrieval}

In code retrieval task, we compare our approach with the following state-of-the-art models as baselines in our evaluation.

\begin{itemize}

\item \textbf{DCS} ~\cite{DCS} is a deep code search model which uses two deep neural networks to encode source code and natural language description into vector representation, and then uses a cosine similarity function to calculate their similarity.

\item \textbf{CoaCor} ~\cite{CoaCor} is a code annotation model trained to generate a natural language annotation which represents the semantics of a given code snippet. Then it projects the annotation and candidate code snippets into a high-dimensional vector space to calculate cosine similarity, thus fulfilling the code retrieval task.

\end{itemize}

\subsubsection{Code Summarization}

In the code summarization task, we compare our approach with the following baselines.

\begin{itemize}

\item \textbf{CODE-NN} ~\cite{CodeNN} is an end-to-end code summarization approach. This approach uses LSTM as the decoder and applies an attention mechanism in each decoding step.

\item \textbf{Seq2Seq} is a basic encoder-decoder model. We choose Bi-LSTM as the RNN layer for the encoder and LSTM for the decoder, and the attention mechanism ~\cite{CodeNN} is applied in each decoding step.

\item \textbf{CoaCor} ~\cite{CoaCor} is a code annotation model, but the annotation can also be considered as the summary of code snippets.
%from others

\end{itemize}

\subsection{CO3 Variants}

In our approach, we introduce an additional code generation task, adopt dual learning to improve the dual tasks of code generation and code summarization, and finally train them with code retrieval task with multi-task learning. To evaluate the effect of the code generation task, dual learning and multi-task learning, we will perform ablation study on the following variants of our model CO3. Note that the architecture of CO3 is extremely simple. Though we model three tasks simultaneously, we actually have only two LSTM instances. The variants mainly involve different combination of loss functions.

\begin{itemize}
\item CO3: When we mention CO3 in our experiment results, it refers to the full model described in Section 3. Loss functions of CO3 consist of ${L}_{cs}$, ${L}_{cg}$, ${L}_{cr}$ and ${L}_{dual}$.

\item CO3($-$Dual Learning)-1: This is a CO3-variant that removes the regularization term of dual learning. The loss functions consist of ${L}_{cs}$, ${L}_{cg}$ and ${L}_{cr}$.

\item CO3($-$Dual Learning)-2: This is a CO3-variant that removes the regularization term of dual learning and parameter sharing between code summarization and code generation, modeling them as two independent tasks. The loss functions consist of ${L}_{cs}$, ${L}_{cg}$ and ${L}_{cr}$.

\item CO3($-$Code Generation): This is a CO3-variant that removes the code generation module, which means it has only the code summarization module and the code retrieval module, sharing only the code encoder. The loss functions consist of ${L}_{cs}$ and ${L}_{cr}$.

\item DCS: This is a CO3-variant that has only code retrieval module, which is a simplified DCS~\cite{DCS}. The loss function is ${L}_{cr}$.

\end{itemize}

\section{Experiments Results}

\subsection{Performance of Code Retrieval}

To evaluate the performance of the code retrieval task, we use DCS and CoaCor as baselines, in which DCS~\cite{DCS} is a very competitive code retrieval models from software engineering community and CoaCor~\cite{CoaCor} is a state-of-the-art model proposed recently. As shown in Table~\ref{tab:CR_baselines}, among the three models, CO3 achieves the highest score across all metrics for both SQL and Python. Since CoaCor uses ranking metrics for retrieval as reward to producing retrieval-friendly code summary and further assembles two diverse retrieval models, CoaCor outperforms DCS by a large margin. Despite the simplicity of CO3, it outperforms CoaCor by nearly 0.01 for SQL in terms of both MRR and NDGG. The performance margin between CoaCor and CO3 is even larger for Python, which speaks to the superiority of CO3.

In addition to achieving a new state-of-the-art model for code retrieval, more importantly, we do not sacrifice the BLEU score of code summarization task, as evidenced by the following experiment results. 

\begin{table}
	\caption{Code Retrieval Results of CO3 and Baselines. We appiyed T-test on the results between Our Model and baselines, and all p-values are smaller than 0.05, which means that the improvement is stable and statistically significant.}
	\label{tab:CR_baselines}
	\begin{tabular}{crrrrrr}
		\hline
		\multirow{2}*{Model} & \multicolumn{2}{c}{SQL} & \multicolumn{2}{c}{Python} \\
		\cline{2-5}
		 & MRR & NDCG & MRR & NDCG \\
		\hline
		DCS     & 0.522 & 0.629 & 0.617 & 0.705 \\
		CoaCor  & 0.576 & 0.670 & 0.636  & 0.721 \\
		Our Model & \textbf{0.585} & \textbf{0.679} & \textbf{0.682} & \textbf{0.756} \\
		\hline
	\end{tabular}
\end{table}

\subsection{Performance of Code Summarization}

To evaluate the performance of code summarization task, we use CODE-NN and a strong Seq2Seq model equipped with attention mechanism as baselines. Results in Table~\ref{tab:CS_baselines} show that our model outperforms the two baselines. CO3 is built upon the Seq2Seq model, and we find the introduction of code generation task and the adoption of dual learning and multi-task learning improve the performance by more than $15\%$ in terms of BLEU4 and METEOR, which verifies the effectiveness of our approach. How each of these design choices contributes to the improvement will analysed in Sections 5.4 through 5.6.

We also present CoaCor's scores for code summarization when it achieves the best performance for code retrieval. As we can see, for both languages, CO3 outperforms CoaCor in both BLEU and METEOR by a very large margin, even by more than $100\%$. The reason is that CoaCor prefers to generate code summary with longer length, which contains more conceptual words for code retrieval and thus weakens human readability. In contrast, CO3 generates code summaries with style more in line with human-written queries in training data. 

\subsection{Balance between Code Retrieval and Code Summarization}

\begin{figure}[ht]
	\centering
	\includegraphics[width=\linewidth]{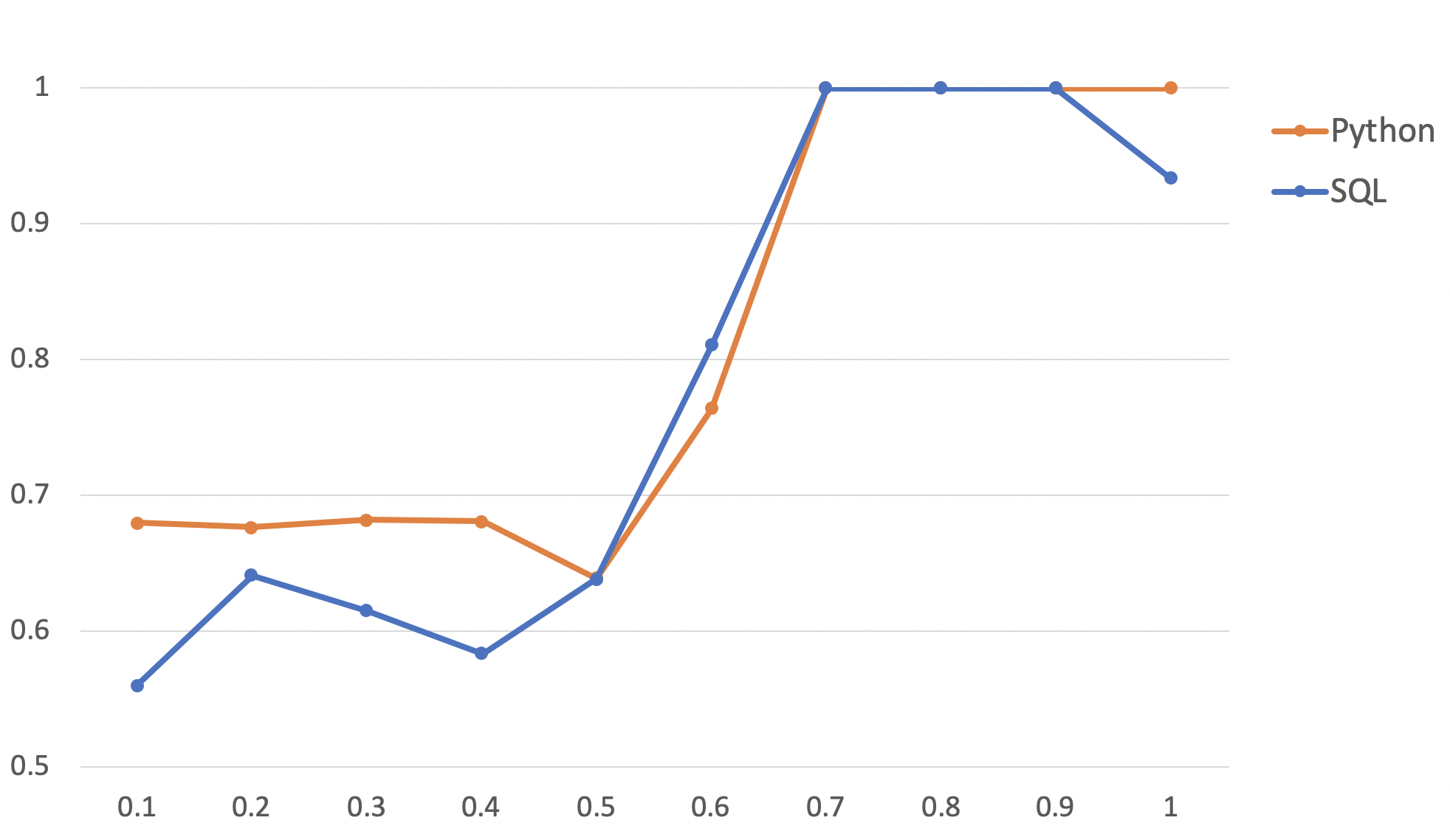}
	\caption{MRR scores for data in the test set of SQL and Python grouped by BLEU score. }
	\label{fig:BLEU_MRR}
\end{figure}

We have shown that our model can significantly improve the results of code retrieval task over the-state-of-art models, as well as achieve competitive performance in terms of BLEU score for code summarization task. To explain why CO3 can balance code retrieval and code summarization better, we divided the test set of SQL dataset and Python dataset into 10 groups according to the BLEU score (within [0, 1]) with an interval of 0.1. Then we calculated the average MRR score for all samples in each group. The results are presented in Figure~\ref{fig:BLEU_MRR}. We can see for both the SQL dataset and Python dataset, as the BLEU score increases, the MRR score also increases. Note that the data with BLEU score over 0.9 are very sparse, and one outlier for the SQL dataset (MRR 0.3 -- BLUE 1.0) affects the average score and causes the MRR plot to go down towards the right end.

The correlation between MRR and BLUE scores corroborates our argument in Section \ref{sec:introduction} that generating summaries close to human-provided queries is naturally fit to code retrieval, and explains why CO3 can excel at code retrieval and code summarization at the same time.

\begin{table}
	\caption{Code Summarization Results of CO3 and Baselines}
	\label{tab:CS_baselines}
	\begin{tabular}{crrrr}
		\hline
		\multirow{2}*{Model} & \multicolumn{2}{c}{SQL} & \multicolumn{2}{c}{Python} \\
		\cline{2-5}
		 & BLEU4 & METEOR & BLEU4 & METEOR \\
		\hline
		CODE-NN   & 0.083 & 0.041 & 0.092 & 0.053 \\
		Seq2Seq   & 0.104 & 0.067 & 0.106 & 0.062 \\
		CoaCor    & 0.067 & 0.029  & 0.078 & 0.034 \\
		Our Model & \textbf{0.119} & \textbf{0.087} & \textbf{0.124} & \textbf{0.085} \\
		\hline
	\end{tabular}
\end{table}

\subsection{Effect of Code Generation Task}

To investigate the effect of code generation task, we first conducted experiments on CO3($-$Code Generation), which consists of only the code summarization module and the code retrieval module. For code retrieval, we find the performance degrades significantly when code generation task is removed. For example, as shown in Table~\ref{tab:CR_variant}, compared with the full CO3 model, the MRR scores of CO3($-$Code Generation) decreased 0.037 and 0.028 for SQL and Python, respectively. For code summarization, the result is similar. As shown in Table~\ref{tab:CS_variant}, the BLEU scores of CO3($-$Code Generation) decreased 0.030 and 0.015 for SQL and Python, respectively. 

Meanwhile, From Table~\ref{tab:CR_variant} and Table~\ref{tab:CS_variant}, we find CO3($-$Dual Learning)-1 (with LSTM parameters shared) and CO3($-$Dual Learning)-2 (with independent LSTM parameters) always outperform CO3($-$Code Generation). So we conclude that the code generation task can be helpful even without dual learning. 

These two observations indicate that the code generation task is beneficial to capturing more accurate semantics of code and the paired natural language text, serving as fundamental component in our architecture.

We also have an interesting observation that the performances of CO3($-$Dual Learning)-1 and CO3($-$Dual Learning)-2 are basically the same, which indicates that whether sharing LSTM parameters for dual learning or not does not have too much effect on the performance. However, with parameter sharing, CO3($-$Dual Learning)-1 reduces nearly half of the parameters of CO3($-$Dual Learning)-2, which makes training more efficient.

\subsection{Effect of Dual Learning}

To investigate the effect of dual learning, we compare the performance of CO3($-$Dual learning) with the full CO3 model. From Table~\ref{tab:CR_variant} and Table~\ref{tab:CS_variant}, we can observe that for both code retrieval and code summarization, full CO3 outperforms CO3($-$Dual learning) for both SQL and Python. Thus, we can conclude that dual learning can help generate more accurate semantic embeddings used by decoders to generate code or text, which is also beneficial to the code retrieval module. 

Compared with code retrieval, the improvement caused by dual learning for code summarization is larger. The reason is that the regularization term of dual learning mainly imposes a direct constraint on the joint probability of the two dual tasks: code summarization and code generation. It is conceivable that dual learning can help generate better code summaries, and the regularization term of dual learning is effective.

\begin{table}
	\caption{Code Retrieval Results of CO3 Variant}
	\label{tab:CR_variant}
	\begin{tabular}{crrrrrr}
		\hline
		\multirow{2}*{Model} & \multicolumn{2}{c}{SQL} & \multicolumn{2}{c}{Python} \\
		\cline{2-5}
		 & MRR & NDCG & MRR & NDCG \\
		\hline
		CO3 & \textbf{0.585} & \textbf{0.679} & \textbf{0.682} & \textbf{0.756} \\
		CO3($-$Dual Learning)-1 & 0.583 & 0.678 & 0.660 & 0.740 \\
		CO3($-$Dual Learning)-2           & 0.581 & 0.675 & 0.664 & 0.742 \\
		CO3($-$Code Generation)       & 0.548 & 0.650 & 0.654 & 0.734 \\
		DCS                         & 0.522 & 0.629 & 0.617 & 0.705 \\
		\hline
	\end{tabular}
\end{table}

\begin{table}
	\caption{Code Summarization Results of CO3 Variant}
	\label{tab:CS_variant}
	\begin{tabular}{crrrrrr}
		\hline
		\multirow{2}*{Model} & \multicolumn{2}{c}{SQL} & \multicolumn{2}{c}{Python} \\
		\cline{2-5}
		& BLEU4 & METEOR & BLEU4 & METEOR \\
		\hline
		CO3 & 0.119 & 0.087 & 0.124 & 0.085 \\
		CO3($-$Dual Learning)-1 & 0.100 & 0.061 & 0.117 & 0.081 \\
		CO3($-$Dual Learning)-2           & 0.102 & 0.061 & 0.111 & 0.082 \\
		CO3($-$Code Generation)       & 0.089 & 0.051 & 0.109 & 0.080 \\
		\hline
	\end{tabular}
\end{table}

\subsection{Effect of Multi-task Learning}

To study whether code retrieval enhances code summarization, we removed the code retrieval module from CO3 and found that it did not affect the performance of CO3 much.

And for code retrieval, we compare CO3 with an individual code retrieval module, a simplified DCS. As shown in Table~\ref{tab:CR_variant}, since CO3 trains the code retrieval model with the dual Seq2Seq model of code summarization/generation, it achieves an improvement of around $10\%$ in terms of MRR and NDCG compared to DCS. In this multi-task learning architecture, the dual Seq2Seq model can be deemed as an auxiliary task to provide inductive bias, which makes the model more focused on those hypotheses that can explain both the dual Seq2Seq model and code retrieval at the same time, and consequently improve the performance and generalization of code retrieval.

\subsection{Case Study}

\begin{figure*}[ht]
	\centering
	\includegraphics[width=\linewidth]{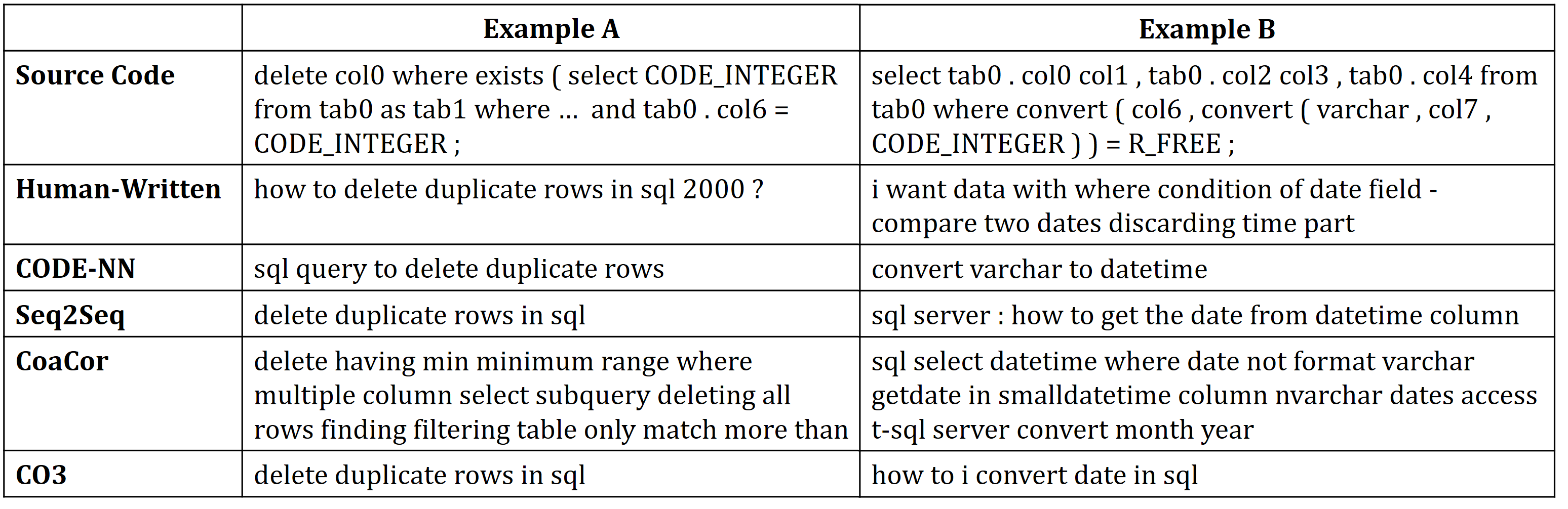}
	\caption{Examples of Code and Generated Code Summaries.}
	\label{fig:CS_examples}
\end{figure*}

\begin{table}
	\caption{Rank of Code Retrieval for Examples.}
	\label{tab:CR_examples}
	\begin{tabular}{crrrrrr}
		\hline
		Model & Example(a) & Example(b) \\
		\hline
		DCS & 3 & 23 \\
		CoaCor & 13 & 9 \\
		CO3 & 1 & 1 \\
		\hline
	\end{tabular}
\end{table}

\begin{figure}[ht]
	\centering
	\includegraphics[width=\linewidth]{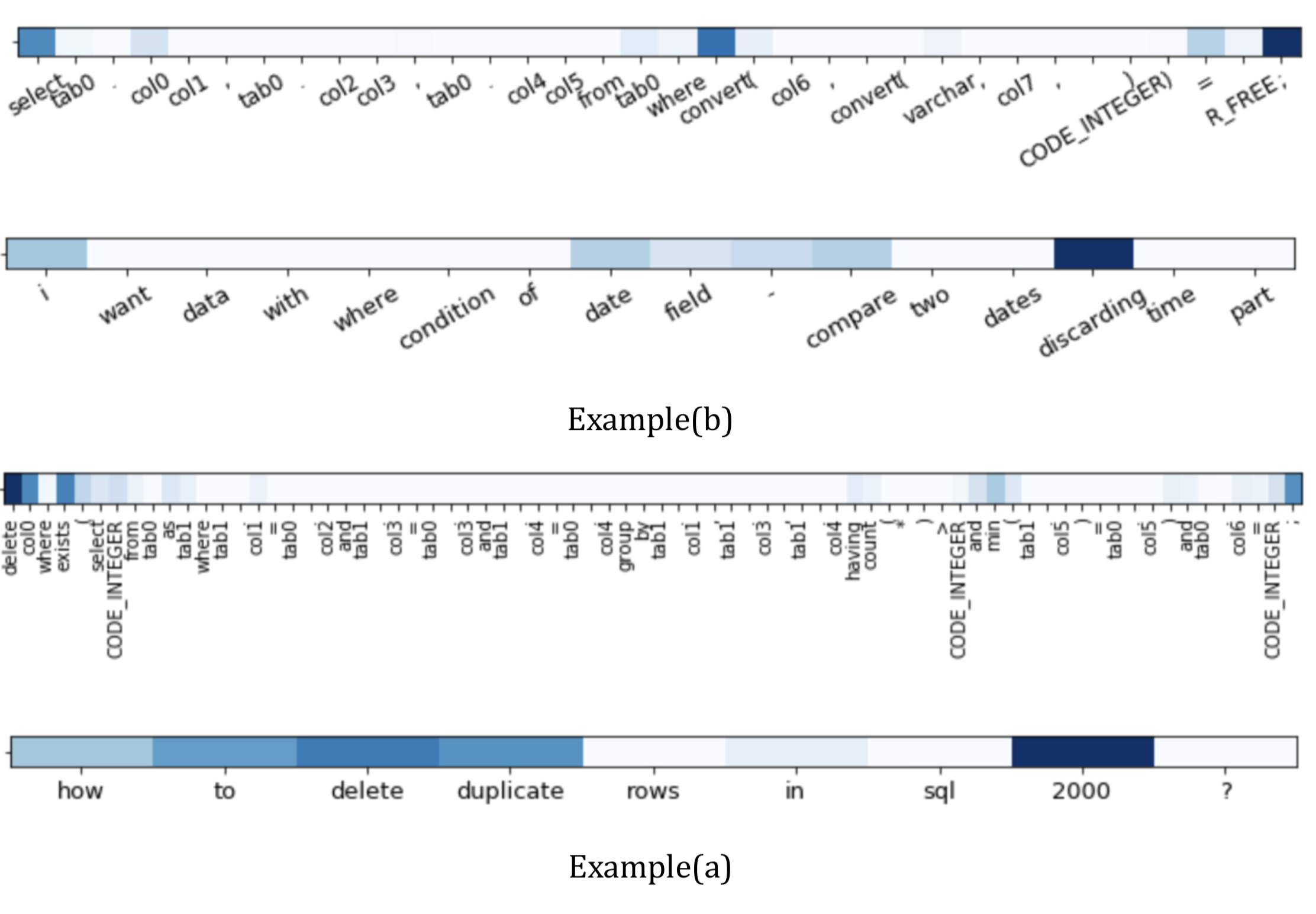}
	\caption{Heatmap of Query and Code Semantic Representation in Max-Pooling Process.}
	\label{fig:heatmap_examples}
\end{figure}

To perform qualitative analysis, we list two examples in Figure~\ref{fig:CS_examples}, with summaries provided by human beings (the titles) and generated by different code summarization models. Their ranking results of each code retrieval model are summarized in Table~\ref{tab:CR_examples}. The examples show that our model not only achieves the best rank in code retrieval task, but also generates a human-readable summary. This indicates that CO3 can capture the inner association between human-provided text and code more accurately, which is consistent with previous quantitative analysis. More details are explained below.

In the first example, all models generated a clear, coherent and informative summary except CoaCor. Although CoaCor did find the keyword ``delete``, it failed to extract other keywords like ``duplicate``. Though the reinforcement learning mechanism can make the generated summaries more friendly to code retrieval, these summaries lost some human readability. In contrast, our model can find more key information while conserving the fluency and naturalness of the generated text. CO3 can easily find the corresponding code snippets with the help of more accurate semantic embeddings, which can be used to generate keyword like ``delete`` and ``duplicate,`` as shown in the code summary generated by CO3. This again shows that a better code summary has the potential to lead to a better performance in code retrieval.

In the second example, due to the more complex logic of the query, none of the code summarization models were able to achieve a good result. Compared with the human-written summary, all the generated summaries lost the key information about ``compare`` and ``discarding.`` However, even in this complex scenario, CO3 did successfully find the corresponding code snippet. In contrast, CoaCor could not locate the correct code snippet, resulting in a target rank of 23, as shown in Table~\ref{tab:CR_examples}. The main reason behind this is that CO3 can capture the intrinsic connection among tasks of code summarization, code generation and code retrieval better through an end-to-end model, thanks to dual learning and multi-task learning. More specifically, the three tasks collaborate to generate more accurate and richer semantic representations of both natural language and source code. These representations are vectors in a continuous semantic space, and contain most of the key information. The discrete words generated by sampling from continuous space may lose some key information, but those continuous representations can still maintain such information and thus assist down-stream tasks such as code retrieval better.

We further transform the representations into a heat map of the query, and find that the words ``compare`` and ``discarding`` are assigned with more weights, as shown in Figure ~\ref{fig:heatmap_examples}. The weight of a word is calculated as the number of times it is selected as the maximum value when performing max-pooling in the scorer of the code retrieval module. That explains why CO3 can retrieve the target code snippets that contains tokens of ``=`` and ``convert`` since in SQL, we often need to use ``=`` to ``compare`` two values and use ``convert`` to ``discarding time part.`` However, CoaCor uses generated summary consisting of discrete words to feed into the code retrieval module in a pipeline manner, which can lead to loss of key information (e.g. ``compare`` and ``discarding``) when sampling from continuous space. Error propagation is a common issue in pipeline-based approaches. This example fully demonstrates that out model can establish better association between natural language and programming language, showing the effectiveness of our end-to-end model.

\section{Related Work}

\subsection{Code Retrieval}

As introduced in previous sections, code retrieval has been studied widely with information retrieval methods~\cite{HaiducBMOLM13,HillRFM14,KeivanlooRZ14,LuSWLD15,VinayakaraoSPJJ17} and recent deep learning methods~\cite{AllamanisTGW15,DCS,CodeNN}. Chen et al. ~\cite{BVAE} used VAEs to model both source code and natural language. Two VAEs are trained jointly to reconstruct their inputs as much as possible with regularization that captures the closeness between the latent variables of code and description, which will be used for measuring similarity. Similarly, Yao et al. \cite{CoaCor} constructed a neural network-based code annotation model to describe the functionality of an entire code snippet. It produces meaningful words that can be used for code retrieval where these words and a natural language query are projected into a vector space to measure the cosine similarity between them. 

Like some of these efforts, our code retrieval model directly encodes the query and code, and projects them into high-dimensional vector space. But we also use dual learning~\cite{DSL} and multi-task learning to catch more intrinsic and precise representations of query and code, which improve the performance significantly.

\subsection{Code Summarization}

The existing works for code summarization can be mainly categorized as traditional approaches based on information retrieval~\cite{Eddy2013Evaluating}, code keywords~\cite{Moreno2013Automatic,sridhara2010towards}, statistical language models~\cite{Dana2013Natural}, and deep learning based approaches~\cite{Allamanis2016A,CodeNN,SBT}. In~\cite{Eddy2013Evaluating}, the authors generated code summarization by searching for similar codes. Sridhara et al.~\cite{sridhara2010towards} generated comments according to the keywords extracted from code. Movshovitz-Attias et al.~\cite{Dana2013Natural} predicted comments from Java source files using topic models and n-grams. Allamanis et al. ~\cite{Allamanis2016A} proposed an attention-based neural network to summarize source code into method name-like summaries. They employed convolution on the source code tokens to detect local time-invariant and long-range topical attention features. Iyer et al.~\cite{CodeNN} proposed an attention-based Recurrent Neural Network (RNN) model, which aligns the words in comments with individual code tokens directly by an attention component. The code summarization task can be modeled as machine translation problem, so some models based on Seq2Seq paradigm~\cite{Sutskever2014Sequence} were proposed. Hu et al.~\cite{SBT} proposed a structure-based traversal (SBT) algorithm in the encoder to flatten an AST and link the tokens in the source code  with their AST node types. 

Different from previous deep learning based works that design better features or more sophisticated network structures, this paper introduces a dual task of code generation to improve the performance of code summarization.

\subsection{Dual Learning}

For multi-tasks learning scenarios where there are two dual tasks of each other, He et al.~\cite{HeXQWYLM16} proposed a new framework to utilize the duality, named dual learning. According to their motivations, dual learning can be applied in two cases: 1) As semi-supervised or unsupervised method to deal with lack of data. 2) as supervised method to regularize the models. For example, ~\cite{DSL} proposed to use dual learning on machine translation with unsupervised data, and leveraged the duality via mutual translation of two languages, such as translating English to Chinese and translating Chinese to English. Later, ~\cite{WangXZBQLL18} utilized this mechanism with transfer learning to transfer the knowledge of duality on dual machine translation among three languages. Also for machine translation, ~\cite{HeXQWYLM16} proposed a supervised learning, called dual supervised learning, resulting in a remarkable improvement on dual tasks involving two languages. Recently, more and more researchers start to apply this mechanism to other tasks. ~\cite{SuHC19} use dual learning directly in natural language understand (NLU) and generation (NLG), where the input of NLU is a natural language sentence, and the input of NLG is semantic frame. In essence, this paper utilizes the dual learning to improve the performance for two tasks with different representations of input. ~\cite{YeLW19} achieved a better performance via jointly learning and dual learning on both Semantic Parser and Natural Language tasks. 

Therefore, to deal with the problem of different representations of input, and to model the inner connection between dual tasks, we employ dual learning on two related tasks: code summarization and code generation.

\subsection{Multi-Task Learning}

Multi-task learning (MTL) is heavily used in machine learning and natural language processing tasks. MTL aims to help improve the learning of a model by leveraging the domain-specific knowledge contained in the training signals of related tasks~\cite{caruana1997multitask}. Usually, relevance among tasks is learned in two ways in deep neural networks: hard parameter sharing and soft parameter sharing of hidden layer~\cite{ruder2017overview}. 

Hard parameter sharing MTL was first proposed in ~\cite{Caruana1993Multitask}, which shares the hidden layer between all tasks and keeps task-specific output layers. Collobert et al.~\cite{collobert2008unified} described a single convolutional neural network architecture trained jointly on NLP tasks such as part-of-speech tags, chunks, named entity tags, and semantic roles. Zheng et al.~\cite{zheng2018same} proposed a module in which all tasks share the same sentence representation and each task can select the task-specific information from the shared sentence representation with attention mechanism. On the other hand, each task in soft parameter MTL contains its own model and parameters, and the parameters are encouraged to be similar with some regularization. For example, Misra et al.~\cite{misra2016cross} connected two separate networks in a soft parameter sharing way. Then the model leverages a unit called cross-stitch to determine how to combine the knowledge learned in other related tasks as task-specific networks. 

Here, we use hard parameter sharing as our multi-task learning method. Our approach not only uses hard parameter sharing, but also adds a regularization term of duality, which resembles soft parameter sharing.

\section{Conclusion}

We have presented an end-to-end model named CO3 for code retrieval and code summarization. CO3 leverages code generation to bridge programming language and natural language better via dual learning and multi-task learning. Though involving three tasks, CO3 has a simple yet effective architecture, which consists of only two LSTM instances. Compared with previous models which process code retrieval and code summarization in an independent or pipeline manner, CO3 can better capture the intrinsic connection between these tasks, so that it not only improves the results of code retrieval over the state of the art, but also balances the performance of the two tasks much better.  

In the future, we plan to further explore the interaction among the ranking loss for code retrieval, the maximal likelihood estimation objective of text generation, and the dual regularization.

%%
%% The next two lines define the bibliography style to be used, and
%% the bibliography file.
\bibliographystyle{ACM-Reference-Format}
\bibliography{references}

%%% -*-BibTeX-*-
%%% Do NOT edit. File created by BibTeX with style
%%% ACM-Reference-Format-Journals [18-Jan-2012].

\begin{thebibliography}{38}

%%% ====================================================================
%%% NOTE TO THE USER: you can override these defaults by providing
%%% customized versions of any of these macros before the \bibliography
%%% command.  Each of them MUST provide its own final punctuation,
%%% except for \shownote{}, \showDOI{}, and \showURL{}.  The latter two
%%% do not use final punctuation, in order to avoid confusing it with
%%% the Web address.
%%%
%%% To suppress output of a particular field, define its macro to expand
%%% to an empty string, or better, \unskip, like this:
%%%
%%% \newcommand{\showDOI}[1]{\unskip}   % LaTeX syntax
%%%
%%% \def \showDOI #1{\unskip}           % plain TeX syntax
%%%
%%% ====================================================================

\ifx \showCODEN    \undefined \def \showCODEN     #1{\unskip}     \fi
\ifx \showDOI      \undefined \def \showDOI       #1{#1}\fi
\ifx \showISBNx    \undefined \def \showISBNx     #1{\unskip}     \fi
\ifx \showISBNxiii \undefined \def \showISBNxiii  #1{\unskip}     \fi
\ifx \showISSN     \undefined \def \showISSN      #1{\unskip}     \fi
\ifx \showLCCN     \undefined \def \showLCCN      #1{\unskip}     \fi
\ifx \shownote     \undefined \def \shownote      #1{#1}          \fi
\ifx \showarticletitle \undefined \def \showarticletitle #1{#1}   \fi
\ifx \showURL      \undefined \def \showURL       {\relax}        \fi
% The following commands are used for tagged output and should be
% invisible to TeX
\providecommand\bibfield[2]{#2}
\providecommand\bibinfo[2]{#2}
\providecommand\natexlab[1]{#1}
\providecommand\showeprint[2][]{arXiv:#2}

\bibitem[\protect\citeauthoryear{Allamanis, Barr, Devanbu, and
  Sutton}{Allamanis et~al\mbox{.}}{2018}]%
        {AllamanisBDS18}
\bibfield{author}{\bibinfo{person}{Miltiadis Allamanis},
  \bibinfo{person}{Earl~T. Barr}, \bibinfo{person}{Premkumar~T. Devanbu}, {and}
  \bibinfo{person}{Charles~A. Sutton}.} \bibinfo{year}{2018}\natexlab{}.
\newblock \showarticletitle{A Survey of Machine Learning for Big Code and
  Naturalness}.
\newblock \bibinfo{journal}{\emph{{ACM} Comput. Surv.}} \bibinfo{volume}{51},
  \bibinfo{number}{4} (\bibinfo{year}{2018}), \bibinfo{pages}{81:1--81:37}.
\newblock
\urldef\tempurl%
\url{https://doi.org/10.1145/3212695}
\showDOI{\tempurl}


\bibitem[\protect\citeauthoryear{Allamanis, Peng, and Sutton}{Allamanis
  et~al\mbox{.}}{2016}]%
        {Allamanis2016A}
\bibfield{author}{\bibinfo{person}{Miltiadis Allamanis}, \bibinfo{person}{Hao
  Peng}, {and} \bibinfo{person}{Charles~A. Sutton}.}
  \bibinfo{year}{2016}\natexlab{}.
\newblock \showarticletitle{A Convolutional Attention Network for Extreme
  Summarization of Source Code}. In \bibinfo{booktitle}{\emph{Proceedings of
  the 33nd International Conference on Machine Learning, {ICML} 2016, New York
  City, NY, USA, June 19-24, 2016}}. \bibinfo{pages}{2091--2100}.
\newblock
\urldef\tempurl%
\url{http://proceedings.mlr.press/v48/allamanis16.html}
\showURL{%
\tempurl}


\bibitem[\protect\citeauthoryear{Allamanis, Tarlow, Gordon, and Wei}{Allamanis
  et~al\mbox{.}}{2015}]%
        {AllamanisTGW15}
\bibfield{author}{\bibinfo{person}{Miltiadis Allamanis},
  \bibinfo{person}{Daniel Tarlow}, \bibinfo{person}{Andrew~D. Gordon}, {and}
  \bibinfo{person}{Yi Wei}.} \bibinfo{year}{2015}\natexlab{}.
\newblock \showarticletitle{Bimodal Modelling of Source Code and Natural
  Language}. In \bibinfo{booktitle}{\emph{Proceedings of the 32nd International
  Conference on Machine Learning, {ICML} 2015, Lille, France, 6-11 July 2015}}.
  \bibinfo{pages}{2123--2132}.
\newblock
\urldef\tempurl%
\url{http://proceedings.mlr.press/v37/allamanis15.html}
\showURL{%
\tempurl}


\bibitem[\protect\citeauthoryear{Alon, Brody, Levy, and Yahav}{Alon
  et~al\mbox{.}}{2019}]%
        {code2seq}
\bibfield{author}{\bibinfo{person}{Uri Alon}, \bibinfo{person}{Shaked Brody},
  \bibinfo{person}{Omer Levy}, {and} \bibinfo{person}{Eran Yahav}.}
  \bibinfo{year}{2019}\natexlab{}.
\newblock \showarticletitle{code2seq: Generating Sequences from Structured
  Representations of Code}. In \bibinfo{booktitle}{\emph{7th International
  Conference on Learning Representations, {ICLR} 2019, New Orleans, LA, USA,
  May 6-9, 2019}}.
\newblock
\urldef\tempurl%
\url{https://openreview.net/forum?id=H1gKYo09tX}
\showURL{%
\tempurl}


\bibitem[\protect\citeauthoryear{Banerjee and Lavie}{Banerjee and
  Lavie}{2005}]%
        {METEOR}
\bibfield{author}{\bibinfo{person}{Satanjeev Banerjee} {and}
  \bibinfo{person}{Alon Lavie}.} \bibinfo{year}{2005}\natexlab{}.
\newblock \showarticletitle{{METEOR:} An Automatic Metric for {MT} Evaluation
  with Improved Correlation with Human Judgments}. In
  \bibinfo{booktitle}{\emph{Proceedings of the Workshop on Intrinsic and
  Extrinsic Evaluation Measures for Machine Translation and/or
  Summarization@ACL 2005, Ann Arbor, Michigan, USA, June 29, 2005}}.
  \bibinfo{pages}{65--72}.
\newblock
\urldef\tempurl%
\url{https://aclanthology.info/papers/W05-0909/w05-0909}
\showURL{%
\tempurl}


\bibitem[\protect\citeauthoryear{Caruana}{Caruana}{1997}]%
        {caruana1997multitask}
\bibfield{author}{\bibinfo{person}{R Caruana}.}
  \bibinfo{year}{1997}\natexlab{}.
\newblock \showarticletitle{Multitask learning: A knowledge-based source of
  inductive bias. Machine Learning}.
\newblock  (\bibinfo{year}{1997}).
\newblock


\bibitem[\protect\citeauthoryear{Caruana}{Caruana}{1993}]%
        {Caruana1993Multitask}
\bibfield{author}{\bibinfo{person}{Richard~A. Caruana}.}
  \bibinfo{year}{1993}\natexlab{}.
\newblock \showarticletitle{Multitask Learning: A Knowledge-Based Source of
  Inductive Bias}.
\newblock \bibinfo{journal}{\emph{Machine Learning Proceedings}}
  \bibinfo{volume}{10}, \bibinfo{number}{1} (\bibinfo{year}{1993}),
  \bibinfo{pages}{41--48}.
\newblock


\bibitem[\protect\citeauthoryear{Chen and Zhou}{Chen and Zhou}{2018}]%
        {BVAE}
\bibfield{author}{\bibinfo{person}{Qingying Chen} {and}
  \bibinfo{person}{Minghui Zhou}.} \bibinfo{year}{2018}\natexlab{}.
\newblock \showarticletitle{A neural framework for retrieval and summarization
  of source code}. In \bibinfo{booktitle}{\emph{Proceedings of the 33rd
  {ACM/IEEE} International Conference on Automated Software Engineering, {ASE}
  2018, Montpellier, France, September 3-7, 2018}}. \bibinfo{pages}{826--831}.
\newblock
\urldef\tempurl%
\url{https://doi.org/10.1145/3238147.3240471}
\showDOI{\tempurl}


\bibitem[\protect\citeauthoryear{Collobert and Weston}{Collobert and
  Weston}{2008}]%
        {collobert2008unified}
\bibfield{author}{\bibinfo{person}{Ronan Collobert} {and}
  \bibinfo{person}{Jason Weston}.} \bibinfo{year}{2008}\natexlab{}.
\newblock \showarticletitle{A unified architecture for natural language
  processing: Deep neural networks with multitask learning}. In
  \bibinfo{booktitle}{\emph{Proceedings of the 25th international conference on
  Machine learning}}. ACM, \bibinfo{pages}{160--167}.
\newblock


\bibitem[\protect\citeauthoryear{Eddy, Robinson, Kraft, and Carver}{Eddy
  et~al\mbox{.}}{2013}]%
        {Eddy2013Evaluating}
\bibfield{author}{\bibinfo{person}{Brian~P. Eddy}, \bibinfo{person}{Jeffrey~A.
  Robinson}, \bibinfo{person}{Nicholas~A. Kraft}, {and}
  \bibinfo{person}{Jeffrey~C. Carver}.} \bibinfo{year}{2013}\natexlab{}.
\newblock \showarticletitle{Evaluating source code summarization techniques:
  Replication and expansion}. In \bibinfo{booktitle}{\emph{{IEEE} 21st
  International Conference on Program Comprehension, {ICPC} 2013, San
  Francisco, CA, USA, 20-21 May, 2013}}. \bibinfo{pages}{13--22}.
\newblock
\urldef\tempurl%
\url{https://doi.org/10.1109/ICPC.2013.6613829}
\showDOI{\tempurl}


\bibitem[\protect\citeauthoryear{Gu, Zhang, and Kim}{Gu et~al\mbox{.}}{2018}]%
        {DCS}
\bibfield{author}{\bibinfo{person}{Xiaodong Gu}, \bibinfo{person}{Hongyu
  Zhang}, {and} \bibinfo{person}{Sunghun Kim}.}
  \bibinfo{year}{2018}\natexlab{}.
\newblock \showarticletitle{Deep code search}. In
  \bibinfo{booktitle}{\emph{Proceedings of the 40th International Conference on
  Software Engineering, {ICSE} 2018, Gothenburg, Sweden, May 27 - June 03,
  2018}}. \bibinfo{pages}{933--944}.
\newblock
\urldef\tempurl%
\url{https://doi.org/10.1145/3180155.3180167}
\showDOI{\tempurl}


\bibitem[\protect\citeauthoryear{Haiduc, Bavota, Marcus, Oliveto, Lucia, and
  Menzies}{Haiduc et~al\mbox{.}}{2013}]%
        {HaiducBMOLM13}
\bibfield{author}{\bibinfo{person}{Sonia Haiduc}, \bibinfo{person}{Gabriele
  Bavota}, \bibinfo{person}{Andrian Marcus}, \bibinfo{person}{Rocco Oliveto},
  \bibinfo{person}{Andrea~De Lucia}, {and} \bibinfo{person}{Tim Menzies}.}
  \bibinfo{year}{2013}\natexlab{}.
\newblock \showarticletitle{Automatic query reformulations for text retrieval
  in software engineering}. In \bibinfo{booktitle}{\emph{35th International
  Conference on Software Engineering, {ICSE} '13, San Francisco, CA, USA, May
  18-26, 2013}}. \bibinfo{pages}{842--851}.
\newblock
\urldef\tempurl%
\url{https://doi.org/10.1109/ICSE.2013.6606630}
\showDOI{\tempurl}


\bibitem[\protect\citeauthoryear{He, Xia, Qin, Wang, Yu, Liu, and Ma}{He
  et~al\mbox{.}}{2016}]%
        {HeXQWYLM16}
\bibfield{author}{\bibinfo{person}{Di He}, \bibinfo{person}{Yingce Xia},
  \bibinfo{person}{Tao Qin}, \bibinfo{person}{Liwei Wang},
  \bibinfo{person}{Nenghai Yu}, \bibinfo{person}{Tie{-}Yan Liu}, {and}
  \bibinfo{person}{Wei{-}Ying Ma}.} \bibinfo{year}{2016}\natexlab{}.
\newblock \showarticletitle{Dual Learning for Machine Translation}. In
  \bibinfo{booktitle}{\emph{Advances in Neural Information Processing Systems
  29: Annual Conference on Neural Information Processing Systems 2016, December
  5-10, 2016, Barcelona, Spain}}. \bibinfo{pages}{820--828}.
\newblock
\urldef\tempurl%
\url{http://papers.nips.cc/paper/6469-dual-learning-for-machine-translation}
\showURL{%
\tempurl}


\bibitem[\protect\citeauthoryear{Hill, Roldan{-}Vega, Fails, and Mallet}{Hill
  et~al\mbox{.}}{2014}]%
        {HillRFM14}
\bibfield{author}{\bibinfo{person}{Emily Hill}, \bibinfo{person}{Manuel
  Roldan{-}Vega}, \bibinfo{person}{Jerry~Alan Fails}, {and}
  \bibinfo{person}{Greg Mallet}.} \bibinfo{year}{2014}\natexlab{}.
\newblock \showarticletitle{NL-based query refinement and contextualized code
  search results: {A} user study}. In \bibinfo{booktitle}{\emph{2014 Software
  Evolution Week - {IEEE} Conference on Software Maintenance, Reengineering,
  and Reverse Engineering, {CSMR-WCRE} 2014, Antwerp, Belgium, February 3-6,
  2014}}. \bibinfo{pages}{34--43}.
\newblock
\urldef\tempurl%
\url{https://doi.org/10.1109/CSMR-WCRE.2014.6747190}
\showDOI{\tempurl}


\bibitem[\protect\citeauthoryear{Hochreiter and Schmidhuber}{Hochreiter and
  Schmidhuber}{1997}]%
        {LSTM}
\bibfield{author}{\bibinfo{person}{Sepp Hochreiter} {and}
  \bibinfo{person}{J{\"{u}}rgen Schmidhuber}.} \bibinfo{year}{1997}\natexlab{}.
\newblock \showarticletitle{Long Short-Term Memory}.
\newblock \bibinfo{journal}{\emph{Neural Computation}} \bibinfo{volume}{9},
  \bibinfo{number}{8} (\bibinfo{year}{1997}), \bibinfo{pages}{1735--1780}.
\newblock
\urldef\tempurl%
\url{https://doi.org/10.1162/neco.1997.9.8.1735}
\showDOI{\tempurl}


\bibitem[\protect\citeauthoryear{Hu, Li, Xia, Lo, and Jin}{Hu
  et~al\mbox{.}}{2018}]%
        {SBT}
\bibfield{author}{\bibinfo{person}{Xing Hu}, \bibinfo{person}{Ge Li},
  \bibinfo{person}{Xin Xia}, \bibinfo{person}{David Lo}, {and}
  \bibinfo{person}{Zhi Jin}.} \bibinfo{year}{2018}\natexlab{}.
\newblock \showarticletitle{Deep code comment generation}. In
  \bibinfo{booktitle}{\emph{Proceedings of the 26th Conference on Program
  Comprehension, {ICPC} 2018, Gothenburg, Sweden, May 27-28, 2018}}.
  \bibinfo{pages}{200--210}.
\newblock
\urldef\tempurl%
\url{https://doi.org/10.1145/3196321.3196334}
\showDOI{\tempurl}


\bibitem[\protect\citeauthoryear{Iyer, Konstas, Cheung, and Zettlemoyer}{Iyer
  et~al\mbox{.}}{2016}]%
        {CodeNN}
\bibfield{author}{\bibinfo{person}{Srinivasan Iyer}, \bibinfo{person}{Ioannis
  Konstas}, \bibinfo{person}{Alvin Cheung}, {and} \bibinfo{person}{Luke
  Zettlemoyer}.} \bibinfo{year}{2016}\natexlab{}.
\newblock \showarticletitle{Summarizing Source Code using a Neural Attention
  Model}. In \bibinfo{booktitle}{\emph{Proceedings of the 54th Annual Meeting
  of the Association for Computational Linguistics, {ACL} 2016}}.
\newblock
\urldef\tempurl%
\url{https://www.aclweb.org/anthology/P16-1195/}
\showURL{%
\tempurl}


\bibitem[\protect\citeauthoryear{Keivanloo, Rilling, and Zou}{Keivanloo
  et~al\mbox{.}}{2014}]%
        {KeivanlooRZ14}
\bibfield{author}{\bibinfo{person}{Iman Keivanloo}, \bibinfo{person}{Juergen
  Rilling}, {and} \bibinfo{person}{Ying Zou}.} \bibinfo{year}{2014}\natexlab{}.
\newblock \showarticletitle{Spotting working code examples}. In
  \bibinfo{booktitle}{\emph{36th International Conference on Software
  Engineering, {ICSE} '14, Hyderabad, India - May 31 - June 07, 2014}}.
  \bibinfo{pages}{664--675}.
\newblock
\urldef\tempurl%
\url{https://doi.org/10.1145/2568225.2568292}
\showDOI{\tempurl}


\bibitem[\protect\citeauthoryear{Lu, Sun, Wang, Lo, and Duan}{Lu
  et~al\mbox{.}}{2015}]%
        {LuSWLD15}
\bibfield{author}{\bibinfo{person}{Meili Lu}, \bibinfo{person}{Xiaobing Sun},
  \bibinfo{person}{Shaowei Wang}, \bibinfo{person}{David Lo}, {and}
  \bibinfo{person}{Yucong Duan}.} \bibinfo{year}{2015}\natexlab{}.
\newblock \showarticletitle{Query expansion via WordNet for effective code
  search}. In \bibinfo{booktitle}{\emph{22nd {IEEE} International Conference on
  Software Analysis, Evolution, and Reengineering, {SANER} 2015, Montreal, QC,
  Canada, March 2-6, 2015}}. \bibinfo{pages}{545--549}.
\newblock
\urldef\tempurl%
\url{https://doi.org/10.1109/SANER.2015.7081874}
\showDOI{\tempurl}


\bibitem[\protect\citeauthoryear{Misra, Shrivastava, Gupta, and Hebert}{Misra
  et~al\mbox{.}}{2016}]%
        {misra2016cross}
\bibfield{author}{\bibinfo{person}{Ishan Misra}, \bibinfo{person}{Abhinav
  Shrivastava}, \bibinfo{person}{Abhinav Gupta}, {and} \bibinfo{person}{Martial
  Hebert}.} \bibinfo{year}{2016}\natexlab{}.
\newblock \showarticletitle{Cross-stitch networks for multi-task learning}. In
  \bibinfo{booktitle}{\emph{Proceedings of the IEEE Conference on Computer
  Vision and Pattern Recognition}}. \bibinfo{pages}{3994--4003}.
\newblock


\bibitem[\protect\citeauthoryear{Moreno, Aponte, Sridhara, Marcus, Pollock, and
  Vijay-Shanker}{Moreno et~al\mbox{.}}{2013}]%
        {Moreno2013Automatic}
\bibfield{author}{\bibinfo{person}{Laura Moreno}, \bibinfo{person}{Jairo
  Aponte}, \bibinfo{person}{Giriprasad Sridhara}, \bibinfo{person}{Andrian
  Marcus}, \bibinfo{person}{Lori Pollock}, {and} \bibinfo{person}{K.
  Vijay-Shanker}.} \bibinfo{year}{2013}\natexlab{}.
\newblock \showarticletitle{Automatic generation of natural language summaries
  for Java classes}. In \bibinfo{booktitle}{\emph{IEEE International Conference
  on Program Comprehension}}.
\newblock


\bibitem[\protect\citeauthoryear{Movshovitz{-}Attias and
  Cohen}{Movshovitz{-}Attias and Cohen}{2013}]%
        {Dana2013Natural}
\bibfield{author}{\bibinfo{person}{Dana Movshovitz{-}Attias} {and}
  \bibinfo{person}{William~W. Cohen}.} \bibinfo{year}{2013}\natexlab{}.
\newblock \showarticletitle{Natural Language Models for Predicting Programming
  Comments}. In \bibinfo{booktitle}{\emph{Proceedings of the 51st Annual
  Meeting of the Association for Computational Linguistics, {ACL} 2013, 4-9
  August 2013, Sofia, Bulgaria, Volume 2: Short Papers}}.
  \bibinfo{pages}{35--40}.
\newblock
\urldef\tempurl%
\url{https://www.aclweb.org/anthology/P13-2007/}
\showURL{%
\tempurl}


\bibitem[\protect\citeauthoryear{Overflow}{Overflow}{2019}]%
        {SO}
\bibfield{author}{\bibinfo{person}{Stack Overflow}.}
  \bibinfo{year}{2019}\natexlab{}.
\newblock \bibinfo{booktitle}{\emph{Stack Overflow}}.
\newblock
\urldef\tempurl%
\url{httphttps://stackoverflow.com/}
\showURL{%
\tempurl}


\bibitem[\protect\citeauthoryear{Papineni, Roukos, Ward, and Zhu}{Papineni
  et~al\mbox{.}}{2002}]%
        {BLEU}
\bibfield{author}{\bibinfo{person}{Kishore Papineni}, \bibinfo{person}{Salim
  Roukos}, \bibinfo{person}{Todd Ward}, {and} \bibinfo{person}{Wei{-}Jing
  Zhu}.} \bibinfo{year}{2002}\natexlab{}.
\newblock \showarticletitle{Bleu: a Method for Automatic Evaluation of Machine
  Translation}. In \bibinfo{booktitle}{\emph{Proceedings of the 40th Annual
  Meeting of the Association for Computational Linguistics, July 6-12, 2002,
  Philadelphia, PA, {USA.}}} \bibinfo{pages}{311--318}.
\newblock
\urldef\tempurl%
\url{http://www.aclweb.org/anthology/P02-1040.pdf}
\showURL{%
\tempurl}


\bibitem[\protect\citeauthoryear{Ruder}{Ruder}{2017}]%
        {ruder2017overview}
\bibfield{author}{\bibinfo{person}{Sebastian Ruder}.}
  \bibinfo{year}{2017}\natexlab{}.
\newblock \showarticletitle{An overview of multi-task learning in deep neural
  networks}.
\newblock \bibinfo{journal}{\emph{arXiv preprint arXiv:1706.05098}}
  (\bibinfo{year}{2017}).
\newblock


\bibitem[\protect\citeauthoryear{Sridhara, Hill, Muppaneni, Pollock, and
  Vijay-Shanker}{Sridhara et~al\mbox{.}}{2010}]%
        {sridhara2010towards}
\bibfield{author}{\bibinfo{person}{Giriprasad Sridhara}, \bibinfo{person}{Emily
  Hill}, \bibinfo{person}{Divya Muppaneni}, \bibinfo{person}{Lori Pollock},
  {and} \bibinfo{person}{K Vijay-Shanker}.} \bibinfo{year}{2010}\natexlab{}.
\newblock \showarticletitle{Towards automatically generating summary comments
  for java methods}. In \bibinfo{booktitle}{\emph{Proceedings of the IEEE/ACM
  international conference on Automated software engineering}}. ACM,
  \bibinfo{pages}{43--52}.
\newblock


\bibitem[\protect\citeauthoryear{Su, Huang, and Chen}{Su et~al\mbox{.}}{2019}]%
        {SuHC19}
\bibfield{author}{\bibinfo{person}{Shang{-}Yu Su}, \bibinfo{person}{Chao{-}Wei
  Huang}, {and} \bibinfo{person}{Yun{-}Nung Chen}.}
  \bibinfo{year}{2019}\natexlab{}.
\newblock \showarticletitle{Dual Supervised Learning for Natural Language
  Understanding and Generation}. In \bibinfo{booktitle}{\emph{Proceedings of
  the 57th Conference of the Association for Computational Linguistics, {ACL}
  2019, Florence, Italy, July 28- August 2, 2019, Volume 1: Long Papers}}.
  \bibinfo{pages}{5472--5477}.
\newblock
\urldef\tempurl%
\url{https://www.aclweb.org/anthology/P19-1545/}
\showURL{%
\tempurl}


\bibitem[\protect\citeauthoryear{Sutskever, Vinyals, and V.~Le}{Sutskever
  et~al\mbox{.}}{2014}]%
        {Sutskever2014Sequence}
\bibfield{author}{\bibinfo{person}{Ilya Sutskever}, \bibinfo{person}{Oriol
  Vinyals}, {and} \bibinfo{person}{Quoc V.~Le}.}
  \bibinfo{year}{2014}\natexlab{}.
\newblock \showarticletitle{Sequence to Sequence Learning with Neural
  Networks}.
\newblock \bibinfo{journal}{\emph{Advances in Neural Information Processing
  Systems}}  \bibinfo{volume}{4} (\bibinfo{date}{09} \bibinfo{year}{2014}).
\newblock


\bibitem[\protect\citeauthoryear{Vinayakarao, Sarma, Purandare, Jain, and
  Jain}{Vinayakarao et~al\mbox{.}}{2017}]%
        {VinayakaraoSPJJ17}
\bibfield{author}{\bibinfo{person}{Venkatesh Vinayakarao},
  \bibinfo{person}{Anita Sarma}, \bibinfo{person}{Rahul Purandare},
  \bibinfo{person}{Shuktika Jain}, {and} \bibinfo{person}{Saumya Jain}.}
  \bibinfo{year}{2017}\natexlab{}.
\newblock \showarticletitle{{ANNE:} Improving Source Code Search using Entity
  Retrieval Approach}. In \bibinfo{booktitle}{\emph{Proceedings of the Tenth
  {ACM} International Conference on Web Search and Data Mining, {WSDM} 2017,
  Cambridge, United Kingdom, February 6-10, 2017}}. \bibinfo{pages}{211--220}.
\newblock
\urldef\tempurl%
\url{https://doi.org/10.1145/3018661.3018691}
\showDOI{\tempurl}


\bibitem[\protect\citeauthoryear{Voorhees}{Voorhees}{1999}]%
        {MRR}
\bibfield{author}{\bibinfo{person}{Ellen~M. Voorhees}.}
  \bibinfo{year}{1999}\natexlab{}.
\newblock \showarticletitle{The {TREC-8} Question Answering Track Report}. In
  \bibinfo{booktitle}{\emph{Proceedings of The Eighth Text REtrieval
  Conference, {TREC} 1999, Gaithersburg, Maryland, USA, November 17-19, 1999}}.
\newblock
\urldef\tempurl%
\url{http://trec.nist.gov/pubs/trec8/papers/qa\_report.pdf}
\showURL{%
\tempurl}


\bibitem[\protect\citeauthoryear{Wan, Zhao, Yang, Xu, and Yu}{Wan
  et~al\mbox{.}}{2018}]%
        {Yao2018Improving}
\bibfield{author}{\bibinfo{person}{Yao Wan}, \bibinfo{person}{Zhou Zhao},
  \bibinfo{person}{Min Yang}, \bibinfo{person}{Guandong Xu}, {and}
  \bibinfo{person}{Philip~S. Yu}.} \bibinfo{year}{2018}\natexlab{}.
\newblock \showarticletitle{Improving automatic source code summarization via
  deep reinforcement learning}. In \bibinfo{booktitle}{\emph{the 33rd ACM/IEEE
  International Conference}}.
\newblock


\bibitem[\protect\citeauthoryear{Wang, Xia, Zhao, Bian, Qin, Liu, and Liu}{Wang
  et~al\mbox{.}}{2018}]%
        {WangXZBQLL18}
\bibfield{author}{\bibinfo{person}{Yijun Wang}, \bibinfo{person}{Yingce Xia},
  \bibinfo{person}{Li Zhao}, \bibinfo{person}{Jiang Bian}, \bibinfo{person}{Tao
  Qin}, \bibinfo{person}{Guiquan Liu}, {and} \bibinfo{person}{Tie{-}Yan Liu}.}
  \bibinfo{year}{2018}\natexlab{}.
\newblock \showarticletitle{Dual Transfer Learning for Neural Machine
  Translation with Marginal Distribution Regularization}. In
  \bibinfo{booktitle}{\emph{Proceedings of the Thirty-Second {AAAI} Conference
  on Artificial Intelligence, (AAAI-18), the 30th innovative Applications of
  Artificial Intelligence (IAAI-18), and the 8th {AAAI} Symposium on
  Educational Advances in Artificial Intelligence (EAAI-18), New Orleans,
  Louisiana, USA, February 2-7, 2018}}. \bibinfo{pages}{5553--5560}.
\newblock
\urldef\tempurl%
\url{https://www.aaai.org/ocs/index.php/AAAI/AAAI18/paper/view/17041}
\showURL{%
\tempurl}


\bibitem[\protect\citeauthoryear{Xia, Qin, Chen, Bian, Yu, and Liu}{Xia
  et~al\mbox{.}}{2017}]%
        {DSL}
\bibfield{author}{\bibinfo{person}{Yingce Xia}, \bibinfo{person}{Tao Qin},
  \bibinfo{person}{Wei Chen}, \bibinfo{person}{Jiang Bian},
  \bibinfo{person}{Nenghai Yu}, {and} \bibinfo{person}{Tie{-}Yan Liu}.}
  \bibinfo{year}{2017}\natexlab{}.
\newblock \showarticletitle{Dual Supervised Learning}. In
  \bibinfo{booktitle}{\emph{Proceedings of the 34th International Conference on
  Machine Learning, {ICML} 2017, Sydney, NSW, Australia, 6-11 August 2017}}.
  \bibinfo{pages}{3789--3798}.
\newblock
\urldef\tempurl%
\url{http://proceedings.mlr.press/v70/xia17a.html}
\showURL{%
\tempurl}


\bibitem[\protect\citeauthoryear{Yao, Peddamail, and Sun}{Yao
  et~al\mbox{.}}{2019}]%
        {CoaCor}
\bibfield{author}{\bibinfo{person}{Ziyu Yao},
  \bibinfo{person}{Jayavardhan~Reddy Peddamail}, {and} \bibinfo{person}{Huan
  Sun}.} \bibinfo{year}{2019}\natexlab{}.
\newblock \showarticletitle{CoaCor: Code Annotation for Code Retrieval with
  Reinforcement Learning}. In \bibinfo{booktitle}{\emph{The World Wide Web
  Conference, {WWW} 2019, San Francisco, CA, USA, May 13-17, 2019}}.
  \bibinfo{pages}{2203--2214}.
\newblock
\urldef\tempurl%
\url{https://doi.org/10.1145/3308558.3313632}
\showDOI{\tempurl}


\bibitem[\protect\citeauthoryear{Yao, Weld, Chen, and Sun}{Yao
  et~al\mbox{.}}{2018}]%
        {StaQC}
\bibfield{author}{\bibinfo{person}{Ziyu Yao}, \bibinfo{person}{Daniel~S. Weld},
  \bibinfo{person}{Wei{-}Peng Chen}, {and} \bibinfo{person}{Huan Sun}.}
  \bibinfo{year}{2018}\natexlab{}.
\newblock \showarticletitle{StaQC: {A} Systematically Mined Question-Code
  Dataset from Stack Overflow}. In \bibinfo{booktitle}{\emph{Proceedings of the
  2018 World Wide Web Conference on World Wide Web, {WWW} 2018, Lyon, France,
  April 23-27, 2018}}. \bibinfo{pages}{1693--1703}.
\newblock
\urldef\tempurl%
\url{https://doi.org/10.1145/3178876.3186081}
\showDOI{\tempurl}


\bibitem[\protect\citeauthoryear{Ye, Li, and Wang}{Ye et~al\mbox{.}}{2019}]%
        {YeLW19}
\bibfield{author}{\bibinfo{person}{Hai Ye}, \bibinfo{person}{Wenjie Li}, {and}
  \bibinfo{person}{Lu Wang}.} \bibinfo{year}{2019}\natexlab{}.
\newblock \showarticletitle{Jointly Learning Semantic Parser and Natural
  Language Generator via Dual Information Maximization}. In
  \bibinfo{booktitle}{\emph{Proceedings of the 57th Conference of the
  Association for Computational Linguistics, {ACL} 2019, Florence, Italy, July
  28- August 2, 2019, Volume 1: Long Papers}}. \bibinfo{pages}{2090--2101}.
\newblock
\urldef\tempurl%
\url{https://www.aclweb.org/anthology/P19-1201/}
\showURL{%
\tempurl}


\bibitem[\protect\citeauthoryear{Yining~Wang}{Yining~Wang}{2013}]%
        {NDCG}
\bibfield{author}{\bibinfo{person}{Yuanzhi Li Di He Wei Chen Tie-Yan~Liu.
  Yining~Wang, Liwei~Wang}.} \bibinfo{year}{2013}\natexlab{}.
\newblock \showarticletitle{A Theoretical Analysis of Normalized Discounted
  Cumulative Gain (NDCG) Ranking Measures.}. In \bibinfo{booktitle}{\emph{In
  Proceedings of the 26th Annual Conference on Learning Theory (COLT 2013).}}
\newblock


\bibitem[\protect\citeauthoryear{Zheng, Chen, and Qiu}{Zheng
  et~al\mbox{.}}{2018}]%
        {zheng2018same}
\bibfield{author}{\bibinfo{person}{Renjie Zheng}, \bibinfo{person}{Junkun
  Chen}, {and} \bibinfo{person}{Xipeng Qiu}.} \bibinfo{year}{2018}\natexlab{}.
\newblock \showarticletitle{Same representation, different attentions:
  Shareable sentence representation learning from multiple tasks}.
\newblock \bibinfo{journal}{\emph{arXiv preprint arXiv:1804.08139}}
  (\bibinfo{year}{2018}).
\newblock


\end{thebibliography}

\end{document}